\documentclass[prd,aps,showpacs,nofootinbib,%preprint,
]{revtex4}
\usepackage{amsmath}
\usepackage{amssymb}
\usepackage{amsfonts}
\usepackage{graphicx,bm}
\usepackage{dcolumn}
\usepackage{color,amsxtra}
\usepackage{epsf}
\usepackage{enumerate}
\usepackage{hhline}
\usepackage{array}
\usepackage{tabularx}
\newcommand{\be}{\begin{equation}}
\newcommand{\ee}{\end{equation}}
\newcommand{\bea}{\begin{eqnarray}}
\newcommand{\eea}{\end{eqnarray}}
\newcommand{\beaa}{\begin{eqnarray*}}
\newcommand{\eeaa}{\end{eqnarray*}}

\newcommand{\nn}{\nonumber \\}
\newcommand{\e}{\mathrm{e}}

\def\be{\begin{equation}}
\def\ee{\end{equation}}
\def\bea{\begin{eqnarray}}
\def\eea{\end{eqnarray}}

\def\nn{\nonumber \\}
\def\e{\mathrm{e}}

\begin{document}

\tolerance=5000

\title{Anti-Evaporation of Schwarzschild-de Sitter Black Holes in $F(R)$
gravity}

\author{Shin'ichi Nojiri$^{1,2}$
and Sergei D. Odintsov$^{3,4}$}

\affiliation{ $^1$ Department of Physics, Nagoya University, Nagoya
464-8602, Japan \\
$^2$ Kobayashi-Maskawa Institute for the Origin of Particles and
the Universe, Nagoya University, Nagoya 464-8602, Japan \\
$^3$Instituci\`{o} Catalana de Recerca i Estudis Avan\c{c}ats
(ICREA) and Institut de Ciencies de l'Espai (IEEC-CSIC),Facultad de
Ciencies,C5, Campus UAB, 08193 Barcelona, Spain \\
$^4$ Tomsk State Pedagogical Univ., Tomsk, Russia and Eurasian National Univ.,
Astana, Kazakhstan}

\begin{abstract}
We studied the anti-evaporation of degenerate Schwarzschild-de Sitter
black hole (so-called Nariai space-time) in modified $F(R)$
gravity. The analysis of perturbations of the Nariai black hole is done with
the conclusion that anti-evaporation may occur in such a theory already on
classical level.
For several power-law $F(R)$ gravities which may describe the inflation
and/or dark energy eras we presented the theory parameters bounds for
occurrence of anti-evaporation and conjectured creation of infinite 
number of horizons.
\end{abstract}


\maketitle

It is rather well-known that
the horizon radius of the black hole in the vacuum usually decreases by
the Hawking radiation.
However, Bousso and Hawking \cite{Bousso:1997wi} have observed a
phenomenon where the black hole radius increases by the quantum
correction for specific (degenerate) Nariai black hole \cite{Nariai}.
This phenomenon is called anti-evaporation of black holes.
The Nariai space-time appears in the limit where the radius of the cosmological
horizon coincides with
the radius of the black hole horizon in the Schwarzschild- de-Sitter
space-time. Note that such black holes are primordial ones, not those
which appeared in the star collapse. It is then expected that such
primordial black holes should quickly decay due to the Hawking radiation.
In other words, such black holes are not expected to be observed at current
epoch.
In Ref.~\cite{Bousso:1997wi}, the quantum effects were accounted in $s$-wave
and one-loop approximation, in other words, four-dimensional space-time is
reduced to effective two-dimensional space with dynamical dilaton and
two-dimensional conformal anomaly induced effective action was applied.
After that, in \cite{Nojiri:1998ue}, the anti-evaporation was investigated by
using more general two-dimensional effective action.
However, $s$-wave approximation may be often not reliable.
Hence, in Ref.~\cite{Nojiri:1998ph},
consistent four-dimensional approach with account of the full
four-dimensional conformal anomaly induced effective action was
used for the analysis. It was demonstrated that the anti-evaporation could
occur in the full four-dimensional theory (no $s$-wave reduction on quantum
level).

In this paper, we study the anti-evaporation of the Nariai black holes
in  $F(R)$ gravity, which is a theory describing inflation in the early universe
and/or accelerated expansion in the present
universe (for review of unified inflation-dark energy theories of modified
gravity, see Ref.~\cite{review} while for black holes issue
in modified gravity, see
\cite{Clifton:2011jh,Olmo:2006eh,Briscese:2007cd,delaCruzDombriz:2009et,DeLaurentis:2012st}).
The non-linear structure of $F(R)$ gravity
may play the role of the conformal anomaly
induced effective action which gives the anti-evaporation. If the
anti-evaporation occurs in $F(R)$ gravity, the lifetime of
the primordial black holes could become large and they might survive even
in the present universe. If so they probably may be identified with the
dark matter. From another side, their presence in the current universe may give
the observational evidence for such theories of modified gravity.
We make the analysis of perturbations of the Nariai space in $F(R)$ gravity
and show that anti-evaporation may occur already on classical level.
The explicit examples of modified gravity supporting inflation and leading to
anti-evaporation are presented.

The equation corresponding to the Einstein equation in the $F(R)$ gravity
is given by \footnote{
We use the following convention for the curvatures:
\[
R=g^{\mu\nu}R_{\mu\nu} \, , \quad
R_{\mu\nu} = R^\lambda_{\ \mu\lambda\nu} \, , \quad
R^\lambda_{\ \mu\rho\nu} = -\Gamma^\lambda_{\mu\rho,\nu}
+ \Gamma^\lambda_{\mu\nu,\rho} -
\Gamma^\eta_{\mu\rho}\Gamma^\lambda_{\nu\eta}
+ \Gamma^\eta_{\mu\nu}\Gamma^\lambda_{\rho\eta} \, ,\quad
\Gamma^\eta_{\mu\lambda} = \frac{1}{2}g^{\eta\nu}\left(
g_{\mu\nu,\lambda} + g_{\lambda\nu,\mu} - g_{\mu\lambda,\nu}
\right)\, .
\]
}
\be
\label{JGRG13}
\frac{1}{2}g_{\mu\nu} F(R) - R_{\mu\nu} F'(R) - g_{\mu\nu} \Box F'(R)
+ \nabla_\mu \nabla_\nu F'(R)
= - \frac{\kappa^2}{2}T_{\mathrm{matter}\,\mu\nu}\, .
\ee
Without any matter, assuming that the Ricci tensor is covariantly
constant, that is, $R_{\mu\nu}\propto g_{\mu\nu}$, Eq.~(\ref{JGRG13})
reduces to the algebraic equation:
\be
\label{JGRG16}
0 = 2 F(R) - R F'(R)\, .
\ee
If Eq.~(\ref{JGRG16}) has a solution, the (anti-)de Sitter and/or
Schwarzschild- (anti-)de Sitter space-time,
or Kerr - (anti-)de Sitter space would be exact vacuum
solution depending on the sign of $R$ and the geometry of the space-time.
%is an exact vacuum solution.
Since the Nariai space-time is given by the limit of
the Schwarzschild- (anti-)de Sitter space, the Nariai space-time is also a
solution. The metric of the Nariai space-time is given by
\be
\label{Nr1}
ds^2 = \frac{1}{\Lambda^2 \cosh^2 x} \left( - dt^2 + dx^2 \right)
+ \frac{1}{\Lambda^2}d\Omega^2\, ,\quad
d\Omega^2 \equiv d\theta^2 + \sin^2 \theta d\phi^2\, .
\ee
Here $\Lambda$ is a mass scale and $d\Omega^2$ expresses the metric of
a two dimensional unit sphere.
The scalar curvature is given by $R=R_0\equiv 4\Lambda^2$ and therefore
Eq.~(\ref{JGRG16}) gives
\be
\label{Nr2}
0 = F\left( 4\Lambda^2 \right) - 2 \Lambda^2 F' \left( 4\Lambda^2 \right)\, .
\ee
We now consider the perturbation from the solution describing the Nariai
space-time.
The metric is now assumed to be
\be
\label{Nr3}
ds^2 = \e^{2\rho\left(x,t\right)} \left( - dt^2 + dx^2 \right)
+ \e^{-2 \varphi\left(x,t\right)} d\Omega^2\, .
\ee
The connections and curvatures are given in Appendix.
When $T_{\mathrm{matter}\,\mu\nu}=0$, the $(t,t)$, $(x,x)$, $(t,x)$
$\left( \left(x,t\right) \right)$, and
$(\theta,\theta)$ $\left(\left(\phi,\phi\right)\right)$
components in (\ref{JGRG13}) have the following forms:
\begin{align}
\label{Nr5}
0 = & - \frac{\e^{2\rho}}{2} F(R) - \left( - \ddot\rho + 2 \ddot\varphi
+ \rho'' - 2 {\dot\varphi}^2 - 2 \rho' \varphi' - 2 \dot\rho \dot\varphi
\right) F'(R)
+ \frac{\partial^2 F'(R)}{\partial t^2} - \dot\rho \frac{\partial
F'(R)}{\partial t} - \rho' \frac{\partial F'(R)}{\partial x} \nn
& + \e^{2 \varphi} \left\{ - \frac{\partial}{\partial t}
\left( \e^{-2\varphi} \frac{\partial F'(R)}{\partial t}\right)
+ \frac{\partial}{\partial x}
\left( \e^{-2\varphi} \frac{\partial F'(R)}{\partial x}\right)\right\}
\, ,\nn
0 = & \frac{\e^{2\rho}}{2} F(R) - \left( \ddot\rho + 2 \varphi''
 - \rho'' - 2 {\varphi'}^2 - 2 \rho' \varphi' - 2 \dot\rho \dot\varphi \right)
F'(R)
+ \frac{\partial^2 F'(R)}{\partial x^2} - \dot\rho \frac{\partial
F'(R)}{\partial t} - \rho' \frac{\partial F'(R)}{\partial x} \nn
& - \e^{2 \varphi} \left\{ - \frac{\partial}{\partial t}
\left( \e^{-2\varphi} \frac{\partial F'(R)}{\partial t}\right)
+ \frac{\partial}{\partial x}
\left( \e^{-2\varphi} \frac{\partial F'(R)}{\partial x}\right)\right\}
\, ,\nn
0 = & - \left( 2 {\dot\varphi}' - 2 \varphi' \dot\varphi - 2\rho' \dot\varphi
 -2 \dot\rho \varphi' \right) F'(R)
+ \frac{\partial^2 F'(R)}{\partial t \partial x} - \dot\rho \frac{\partial
F'(R)}{\partial x} - \rho' \frac{\partial F'(R)}{\partial t} \, ,\nn
0 = & \frac{\e^{- 2\varphi}}{2} F(R) - \e^{-2 \left(\rho+\varphi\right)}
\left( - \ddot\varphi + \varphi'' - 2 {\varphi'}^2 + 2 {\dot\varphi}^2 \right)
F'(R) - F'(R) + \e^{-2 \left(\rho+\varphi\right)} \left(
\dot\varphi \frac{\partial F'(R)}{\partial t}
  - \varphi' \frac{\partial F'(R)}{\partial x} \right)\nn
& - \e^{- 2\rho} \left\{ - \frac{\partial}{\partial t}
\left( \e^{-2\varphi} \frac{\partial F'(R)}{\partial t}\right)
+ \frac{\partial}{\partial x}
\left( \e^{-2\varphi} \frac{\partial F'(R)}{\partial x}\right)\right\}\, .
\end{align}
We now consider the following perturbation from the Nariai space-time in
(\ref{Nr1}),
\be
\label{Nr6}
\rho = - \ln \left( \Lambda \cosh x \right) + \delta\rho\, ,\quad
\varphi = \ln \Lambda + \delta \varphi\, .
\ee
Then we find
\be
\label{Nr7}
\delta R = 4 \Lambda^2 \left( - \delta\rho + \delta\varphi \right)
+ \Lambda^2 \cosh^2 x \left( 2 \delta \ddot\rho - 2 \delta \rho''
 - 4 \delta \ddot\varphi + \delta \varphi'' \right) \, .
\ee
The perturbed equations look as:
\begin{align}
\label{Nr8}
0 = & \frac{- F'\left( R_0 \right) + 2 \Lambda^2 F'' \left( R_0 \right)}{2
\Lambda^2 \cosh^2 x} \delta R
 - \frac{F \left( R_0 \right)}{\Lambda^2 \cosh^2 x}\delta\rho
 - F' \left( R_0 \right) \left( - \delta \ddot\rho + 2 \delta \ddot\varphi
+ \delta \rho'' + 2 \tanh x \delta\varphi' \right) \nn
& + \tanh x F'' \left( R_0 \right) \delta R' + F'' \left( R_0 \right)
\delta R''\, ,\nn
0 = & - \frac{- F'\left( R_0 \right) + 2 \Lambda^2 F'' \left( R_0 \right)}{2
\Lambda^2 \cosh^2 x} \delta R
+ \frac{F \left( R_0 \right)}{\Lambda^2 \cosh^2 x}\delta\rho
 - F' \left( R_0 \right) \left( \delta \ddot\rho + 2 \delta \varphi''
 - \delta \rho''
+ 2 \tanh x \delta\varphi' \right) + F'' \left( R_0 \right) \delta \ddot R \nn
& + \tanh x F'' \left( R_0 \right) \delta R' \, ,\nn
0 = & - 2 \left( \delta{\dot\varphi}' + \tanh x \delta \dot\varphi \right)
+ \frac{F'' \left( R_0 \right) }{F' \left( R_0 \right) }\left(
\delta {\dot R}' + \tanh x \delta \dot R\right)\, ,\nn
0 = & - \frac{- F'\left( R_0 \right) + 2 \Lambda^2 F'' \left( R_0 \right)}{2
\Lambda^2 } \delta R - \frac{F \left( R_0 \right)}{\Lambda^2} \delta\varphi
 - \cosh^2 x F' \left( R_0 \right) \left( - \delta \ddot\varphi + \delta
\varphi'' \right)
 - \cosh^2 x F'' \left( R_0 \right) \left( - \delta \ddot R + \delta R''
\right) \, .
\end{align}
By combining the first, second, and fourth equations in (\ref{Nr8}), we obtain
\begin{align}
\label{Nr9}
0 = & \frac{- F'\left( R_0 \right) + 2 \Lambda^2 F'' \left( R_0 \right)}{2
\Lambda^2 \cosh^2 x} \delta R
 - F' \left( R_0 \right) \Box \left( \delta \rho - \delta \varphi
 - \frac{F'' \left( R_0 \right) }{2F'' \left( R_0 \right) } \delta R \right)\, , \nn
0 = & \frac{2\Lambda^2}{\cosh^2 x} \delta \varphi + \Box \left( \delta \rho
+ \frac{F'' \left( R_0 \right) }{2F' \left( R_0 \right) } \delta R \right)\, .
\end{align}
Here
\be
\label{Nr10}
\Box = - \frac{\partial^2}{\partial t^2} + \frac{\partial^2}{\partial x^2}\, .
\ee
The third equation in (\ref{Nr8}) can be integrated to give,
\be
\label{Nr11}
   -2 \delta \varphi + \frac{F'' \left( R_0 \right) }{F' \left( R_0 \right)
}\delta R = C_x (x) + \frac{C_t (t)}{\cosh x}\, .
\ee
Here $C_x(x)$ and $C_t(t)$ are arbitrary functions of $x$ and $t$,
respectively.
By using (\ref{Nr9}) and (\ref{Nr11}),  deleting $\delta \rho$ and
$\delta R$  we obtain
\begin{align}
\label{Nr12}
0 = & \frac{1}{\alpha \cosh^2 x}\left( 2 \left(2\alpha - 1 \right)
\delta\varphi
+ \left(\alpha - 1 \right)  \left( C_x(x) + \frac{C_t(t)}{\cosh x} \right)
\right)
+ \Box \left( 3\delta \varphi + C_x(x) + \frac{C_t(t)}{\cosh x} \right) \, .
\end{align}
Here
\be
\label{Nr12b}
\alpha \equiv \frac{2\Lambda^2 F'' \left(R_0\right)}{F' \left(R_0 \right)}
= \frac{F \left(R_0\right) F'' \left(R_0\right)}{F' \left(R_0 \right)^2}\, .
\ee
Let us show the existence of the solution which expresses the 
anti-evaporation.
For this purpose, as in \cite{Bousso:1997wi,Nojiri:1998ue,Nojiri:1998ph},
we assume $\delta\varphi = \varphi_0 \cosh \omega t \cosh^\beta x$ with
constants $\varphi_0$, $\omega$, and $\beta$ and put $C_x(x) = C_t(t) =0$.
Then, one gets
\be
\label{Nr13}
0=\varphi_0 \cosh \omega t \cosh^\beta x \left\{ \cosh^{-2} x
\left( 4 - \frac{2}{\alpha} - 3\beta^2 + \beta \right) - 3 \omega^2
+ 3 \beta^2 + 2\beta \right)\, ,
\ee
that is
\be
\label{Nr14}
0 = 4 - \frac{2}{\alpha}  - 3\beta^2 + 3\beta\, ,\quad
0 = - 3 \omega^2 + 3 \beta^2 \, .
\ee
Then 
\be
\label{NrR1}
\omega=\pm \beta\, ,\quad
\beta = \frac{1}{2}\left( 1 \pm \sqrt{ \frac{19 \alpha - 8}{3\alpha} } \right)\, .
\ee
If
\be
\label{NrR2}
\alpha<0 \ \mbox{or}\ \alpha> \frac{8}{19}\, ,
\ee
$\beta$ and therefore $\omega$ is real.
Because there is always a solution where the real part of $\omega$ is positive,
the solution corresponding to the Nariai space-time is always unstable.

Now the horizon can be defined by
\be
\label{Nr18}
g^{\mu\nu}\nabla_\mu \varphi \nabla_\nu\varphi = 0\, .
\ee
By assuming $\delta\varphi = \varphi_0 \cosh \omega t \cosh^\beta x$ and
$\omega$ and $\beta$ are real, we obtain
\be
\label{Nr19}
\omega^2 \tanh^2 \omega t = \beta^2 \tanh^2 x\, .
\ee
Because $\omega^2=\beta^2$ from (\ref{Nr14}),
\be
\label{Nr20}
\delta\varphi = \delta\varphi_\mathrm{h} \equiv
\varphi_0 \cosh^2 \beta t \, .
\ee
Eq.~(\ref{Nr3}) shows that $\e^{-\varphi}$ can be regarded as a radius 
coordinate.
Using (\ref{Nr1}), one may now define the radius of the horizon by
$r_\mathrm{h} = \e^{-\delta\varphi_\mathrm{h}}/\Lambda$, then we find
\be
\label{Nr21}
r_\mathrm{h} = \frac{\e^{-\varphi_0 \cosh^2 \beta t}}{\Lambda}\, .
\ee
If $\varphi_0 < 0$, $r_\mathrm{h}$ increases,
which may be regarded as anti-evaporation.
When $\beta$ and $\omega$ are complex,
instead of $\delta\varphi = \varphi_0 \cosh \omega t \cosh^\beta x$,
we obtain the following solution for $\varphi$
\be
\label{Nr26}
\delta\varphi = \Re \left\{ \left( C_+ \e^{\beta t} + C_+ \e^{- \beta t} \right) \e^{\beta x}\right\} \, .
\ee
Here $C_\pm$ are a complex numbers. The notation $\Re$ expresses the real part.
Because the real part of $\beta$ is always positive, $\delta\varphi$ increases unless $C_+=0$ when
$t$ increases and therefore the perturbation grows up, which tells that the
solution corresponding to the Nariai space-time is unstable.

Especially, we may consider the following solution in (\ref{Nr26}) as an example:
\be
\label{NrRRR1}
\delta\varphi = \delta{\tilde \varphi}_\mathrm{h} \equiv
\delta\varphi_0 \left\{ \e^{\frac{t+x}{2}}\left( \cos \frac{\gamma\left(t+x\right)}{2}
 - \frac{1}{\gamma}  \sin \frac{\gamma\left(t+x\right)}{2} \right)
+ \e^{\frac{- t+x}{2}}\left( \cos \frac{\gamma\left(t-x\right)}{2}
+ \frac{1}{\gamma}  \sin \frac{\gamma\left(t-x\right)}{2} \right) \right\}\, ,
\ee
which satisfies an initial condition $\delta \dot \varphi = 0$ when $t=0$.
Here we  wrote $\beta$ as
\be
\label{NrRRR2}
\beta = \frac{1}{2}\left( 1 \pm i \sqrt{ \frac{8 - 19 \alpha}{3\alpha} } \right)
= \frac{1}{2} \left( 1 + i \gamma \right)\, .
\ee
Then Eq.~(\ref{Nr18}), which defines the horizon, has the following form:
\be
\label{NrRRR2B}
0 = \frac{\delta \varphi_0^2}{2} \gamma^2 A^2 \e^x \sin\frac{\gamma\left(t+x\right)}{2}
\sin\frac{\gamma\left(t-x\right)}{2} \, .
\ee
The horizon is given by
\be
\label{NrRRR3}
\mbox{(A)}\ x = - t + \frac{2n \pi}{\gamma}\ \mbox{or}\
\mbox{(B)}\ x = t + \frac{2n \pi}{\gamma}\, .
\ee
Then it seems that infinite number of horizons appears. One may speculate 
that 
the original horizon in the Nariai space-time is separated to the infinite 
number (horizons creation).
On the horizon, we find
\begin{align}
\label{NrRRR4}
\mbox{(A)}&\ \varphi = \delta\varphi_0 (-1)^n \left\{ \e^{\frac{n\pi}{\gamma}}
+ \e^{-t + \frac{n\pi}{\gamma} } \left( \cos \left( \gamma t \right)
+ \frac{1}{\gamma} \sin \left( \gamma t \right) \right)\right\}\, , \\
\label{NrRRR5}
\mbox{(B)}&\ \varphi = \delta\varphi_0 (-1)^n \left\{ \e^{\frac{n\pi}{\gamma}}
+ \e^{t + \frac{n\pi}{\gamma} } \left( \cos \left( \gamma t \right)
 - \frac{1}{\gamma} \sin \left( \gamma t \right) \right)\right\}\, .
\end{align}
Then  the radius of horizon, which is defined by
$r_\mathrm{h} = \e^{-\delta{\tilde\varphi}_\mathrm{h}}/\Lambda$, oscillates, that is,
the evaporation and anti-evaporation are iterated.
Especially in case (B) of (\ref{NrRRR5}), the amplitude becomes larger and larger.

As an example, we consider the following model:
\be
\label{Nr31}
F(R)= \frac{R}{2\kappa^2} + f_0 M^{4-2n} R^n\, .
\ee
Here $f_0$ is dimensionless constant and $M$ is a parameter with a dimension of
mass.
Such theory describes $R^2$-inflation \cite{star} when $n=2$ or specific dark
energy \cite{capo} (for recent review, see \cite{capo1,review}).
Then if $(n-2)f_0>0$, Eq.~(\ref{JGRG16}) gives
\be
\label{Nr32}
R = 4\Lambda^2 = \left\{ 2 \left( n - 2 \right) f_0 \kappa^2 M^{4-2n}
\right\}^{ - \frac{1}{n-1}}\, .
\ee
Using (\ref{Nr12b}), we obtain a very simple result:
\be
\label{Nr33}
\alpha = \frac{n}{4}\, .
\ee
Then the previous results can be rewritten as follows,
\begin{itemize}
\item When $n<0$ or $n> \frac{8}{19}$, anti-evaporation occurs.
\item When $0<n< \frac{8}{19}$, an infinite number of horizons appear and the evaporation and
the anti-evaporation are iterated.
\end{itemize}
We should note that the solution for (\ref{JGRG16}) does not exist when
$n=2$ since the $R^2$-theory does not give exactly the de Sitter inflation.

Since the case $n=2$ does not contribute to Eq.~(\ref{JGRG16}),
we may
%%%%%%%%%%%%%
generalize the model in (\ref{Nr31}) as follows,
%% the following model instead of (\ref{Nr31}):
\be
\label{Nr34}
F(R)= \frac{R}{2\kappa^2} + f_0 M^{4-2n} R^n + f_2 R^2\, .
\ee
Even for the model (\ref{Nr34}) which supports inflation, the solution of
Eq.~(\ref{JGRG16})
is not changed from (\ref{Nr32}) but the parameter $\alpha$ is changed
as
\be
\label{Nr35}
\alpha = \frac{n + 2\xi}{4 \left( 1 + \xi\right)}\, ,\quad
\xi \equiv \frac{f_2}{n-1} \left\{ 2 \left( n - 2
\right)n^2\right\}^{\frac{2-n}{1-n}}
f_0^{\frac{1}{1-n}} M^{\frac{4 - 2n}{1 - n}}\, .
\ee
Here $\alpha$ can be changed by the value of $f_2$.
We now use $\xi$ instead of $f_2$.
Then, one gets
\begin{itemize}
\item When $n>2$ and $- \frac{n}{2}<\xi$, $\xi<\frac{32 - 19n}{6}$, or when
$n<2$ and $\xi<-\frac{n}{2}$ or $\xi>\frac{32 - 19n}{6}$, anti-evaporation occurs.
\item When $n>2$ and $\frac{32 - 19n}{6}<\xi< - \frac{n}{2}$, or when
$n<2$ and $-\frac{n}{2}<\xi<\frac{32 - 19n}{6}$,
an infinite number of horizons appear and the evaporation and
the anti-evaporation are iterated.
\end{itemize}

Thus, we presented a number of $F(R)$ theories which may be used for description
of inflation and/or dark energy eras and which support the anti-evaporation of
the Nariai black hole and conjectured creation of infinite number of 
horizons. In the same way, one can analyze other models, for instance,
those which contain positive and negative powers of curvature as well as other
viable models.

In summary, we proved that modified gravity may lead to anti-evaporation
of the Nariai black hole on classical level, without account of
quantum effects. It is clear that similar effect may be expected in other
theories
of modified gravity, like modified Gauss-Bonnet gravity, string-inspired or
non-local gravity (for review, see \cite{review}).
The corresponding investigation will be done elsewhere.
Of course, the process of anti-evaporation should be investigated in detail,
numerically for specific viable models of modified gravity.
Indeed, in $s$-wave and one-loop approximation it was indicated \cite{RB} that
anti-evaporation may be rather transient effect leading to instabilities of
de Sitter space. The same question should be addressed in modified gravity.
However, even if it is transient effect it may lead to
proliferation\footnote{The opposite effect is quantum annihilation of constant
curvature space \cite{b}.} of the de
Sitter space \cite{B,RB} giving the birth to large or infinite number of
universes, perhaps kind of multiverse (for recent example, see \cite{AV}). From
another side, if anti-evaporation is global (not transient) effect then the
lifetime of primordial black holes may be rather large, so that they survive
until the current epoch. If so then their presence may be used as observational
test in favor of modified gravities supporting the anti-evaporation.
Furthermore, it is known that $F(R)$ gravity may provide
very exotic black holes solutions (for recent example, see \cite{Z}) as well as
different black hole instabilities \cite{myung}.
This indicates that anti-evaporation phenomenon or its generalizations may be
expected for other classes of black holes in modified gravity.

\section*{Acknowledgments.}

We are grateful to R.~Bousso and S.~Hawking for helpful discussion of
related questions.
The work by SN is supported in part by Global COE Program of Nagoya University
(G07) provided by the Ministry of Education, Culture, Sports, Science \&
Technology and by the JSPS Grant-in-Aid for Scientific Research (S) \# 22224003
and (C) \# 23540296.
The work by SDO is supported in part by MICINN (Spain),  project
FIS2010-15640 and by AGAUR (Generalitat de Ca\-ta\-lu\-nya),
contract 2009SGR-994 and by project 2.1839.2011 of Min. of Education and
Science (Russia).
SDO also thanks the Yukawa Institute for Theoretical Physics at Kyoto
University,
where this work was done during the Long-term Workshop YITP-T-12-03
on ``Gravity and Cosmology 2012''.

\appendix

\section{Connections and Curvatures in the space-time (\ref{Nr3})}

For the space-time given in (\ref{Nr3}), the connections and the curvatures are
given by
\begin{align}
\label{Nr4}
& \Gamma^t_{tt}=\Gamma^t_{xx} = \Gamma^x_{tx} = \Gamma^x_{xt} = \dot\rho\,
,\quad
\Gamma^x_{xx} = \Gamma^x_{tt} = \Gamma^t_{tx} = \Gamma^t_{xt} = \rho'\, , \nn
& \Gamma^t_{\phi\phi} = \Gamma^t_{\theta\theta} \sin^2 \theta
= - \dot\varphi \e^{-\left(\rho+\varphi\right)} \sin^2 \theta\, ,\quad
\Gamma^x_{\phi\phi} = \Gamma^x_{\theta\theta} \sin^2 \theta
= \varphi' \e^{-\left(\rho+\varphi\right)} \sin^2 \theta\, ,\nn
& \Gamma^\theta_{t\theta} = \Gamma^\theta_{\theta t}
= \Gamma^\phi_{t\phi} = \Gamma^\phi_{\phi t} = - \dot \varphi\, ,\quad
\Gamma^\theta_{x\theta} = \Gamma^\theta_{\theta x}
= \Gamma^\phi_{x\phi} = \Gamma^\phi_{\phi x} = - \varphi'\, ,\quad
\Gamma^\theta_{\phi\phi} = - \sin\theta \cos\theta\, ,\quad
\Gamma^\phi_{\phi\theta} = \Gamma^\phi_{\theta\phi} = \cot\theta\, ,\nn
& R_{tt} = - \ddot\rho + 2 \ddot\varphi + \rho'' - 2 {\dot\varphi}^2
   - 2\dot\rho \dot\varphi - 2 \rho' \varphi'\, ,\quad
R_{xx} = - \rho'' + \ddot\rho + 2 \varphi'' - 2 {\varphi'}^2 - 2 \dot\rho
\dot\varphi
   - 2 \rho' \varphi' \, ,\nn
& R_{tx} = R_{xt} = 2\dot\varphi' - 2 \varphi' \dot\varphi - 2 \rho'
\dot\varphi
-2 \dot \rho \varphi'\, , \quad
R_{\phi\phi} = R_{\theta\theta} \sin^2 \theta
= \left\{ 1 + \e^{- 2 \left(\rho + \varphi\right)}\left( - \ddot\varphi +
\varphi''
+ 2 {\dot\varphi}^2 - 2 {\varphi'}^2 \right) \right\} \sin^2\theta\, , \nn
& R = \left( 2\ddot\rho - 2 \rho'' - 4 \ddot\varphi + 4 \varphi'' + 6
{\dot\varphi}^2
   - 6 {\varphi'}^2 \right) \e^{-2\rho} + 2 \e^{2\varphi} \, ,\quad
   \mbox{other components} = 0\, .
\end{align}
Here $\dot\ \equiv \partial/\partial t$ and $\ '\equiv \partial/\partial x$.

\end{document}